\documentclass{article}
\usepackage[utf8]{inputenc}

\usepackage[margin=0.75in]{geometry}

\usepackage{url}
\usepackage{color}
\usepackage{animate}
\usepackage{textpos}
\usepackage{amsmath}
\usepackage{tcolorbox}

\usepackage{amsthm,amsmath}

\title{\textbf{Prediction of COVID-19 Disease Progression in India} \\
Under the Effect of National Lockdown}
\author{\textbf{Sourish Das}\\
               Chennai Mathematical Institute, India}
\date{First Version: April 07 2020}

\begin{document}

\maketitle

\begin{abstract}
In this policy paper, we implement the epidemiological SIR to estimate the basic reproduction number $\mathcal{R}_0$ at national and state level.  We also developed the  statistical machine learning model to predict the cases ahead of time. Our analysis indicates that the situation of Punjab ($\mathcal{R}_0\approx 16$) is not good. It requires immediate aggressive attention. We see the $\mathcal{R}_0$ for Madhya Pradesh (3.37) , Maharastra (3.25) and Tamil Nadu (3.09) are more than 3.  The $\mathcal{R}_0$ of Andhra Pradesh (2.96), Delhi (2.82) and West Bengal (2.77) is more than the India's $\mathcal{R}_0=2.75$, as of 04 March, 2020. India's $\mathcal{R}_0=2.75$ (as of 04 March, 2020) is very much comparable to Hubei/China at the early disease progression stage. Our analysis indicates that the early disease progression of India is that of similar to China. Therefore, with lockdown in place, India should expect as many as cases if not more like China. If lockdown works, we should expect less than 66,224 cases by May 01,2020. All data and \texttt{R} code for this paper is available from \url{https://github.com/sourish-cmi/Covid19}
\end{abstract}


\section{Introduction}

The World Health Organization (WHO) declared the outbreak of the novel coronavirus, COVID-19, as a pandemic. It will take twelve to eighteen months to develop the vaccine for the COVID-19, \cite{ferguson2020}. The absence of a vaccine makes the situation worse for the already overstretched Indian health care system. For example, the number of hospital beds, per 1000 population, is less than one, \cite{worldbankdatabase} - it is just one indicator to cite the miserable situation of India's health care system. In the absence of a vaccine, the `social distancing' is the optimal strategy to control the spread of novel coronavirus, \cite{ferguson2020}. Other than social distancing, broad base rapid test and cluster tests are essential to identify those who are infected and isolate them. However, India did not have enough testing capacity as it is reported widely in media, \cite{BBC_report_less_test}. Though, Indian scientists recently developed the affordable testing kit for COVID-19, \cite{BBC_report_Indias_COVID19_test}; India needed a complete overhaul of its health care system in a war footing.  In such a situation, India's Prime Minister Narendra Modi announced an unprecedented three-weeks nationwide lockdown on the 24th March 2020. The purpose of the lockdown is to slow down the spread of the novel coronavirus; so that the Govt can take a multi-prong strategy to add more beds in its network of hospitals, scale up the production of the testing kit of the COVID-19 and personal protection equipment (PPE) for the health workers. In such a grim scenario, the important question for Indian health officials is how many new confirmed cases will be seen and by what time; with the hope that the national lockdown will slow down the spread of the virus; which will buy them time to overhaul of the health care system. However, is lockdown going to provide the necessary slow down of the virus spread? Even if the lockdown helps India to control the spread of the virus, it is not economically sustainable to continue the lockdown further, as large the workforce in India employed in the informal sector as a daily wage laborer. Therefore, in this policy paper, we try to estimate the effect of lockdown and set up a track following which we will know if the lockdown is working!

In this paper, we develop the epidemiological SIR model and statistical machine learning model to predict disease progression in India. We implemented the SIR model to estimate the basic reproduction number $\mathcal{R}_0$ at the national and state level. So that we identify which states require more attention. Then we implement the machine learning model to predict the number of cases ahead of time so Indian administration can be better prepared ahead of time.

In Section (\ref{Sec_database}), we introduce the database, from where the data is downloaded and model is built. In Section (\ref{Sec_methodology}), we present the methodology to analyze and predict the data. In Section (\ref{Sec_Analysis_and_Prediction}) we present our analysis and prediction of the Covid-19 disease progression in India. Section (\ref{Sec_Discussion}) concludes the paper with some policy recommendations.

\section{Data}\label{Sec_database}
In this paper, we used  the following major databases. 
\begin{enumerate}
    \item The data repository for the 2019 Novel Coronavirus by the Johns Hopkins University. The database is available here: \url{https://github.com/CSSEGISandData/COVID-19}
    \item Covid19India is a Crowdsourced open source database for India available from: \url{https://www.covid19india.org/}
    \item Kaggle-Covid-19 in India available from \url{https://www.kaggle.com/sudalairajkumar/covid19-in-india}
\end{enumerate}

\section{Methodology}\label{Sec_methodology}
Legendary statistician Prof George Box, once said
\begin{quote}
``All models are wrong, but some are useful", see \cite{Box1976}.
\end{quote}
Keeping this in our mind, here in this paper, we take a model agnostic two-prong approach. One is to understand the severity of the ground situation; and the second is the prediction, which will help the health officials to make the plans accordingly.  The epidemic models for infectious disease yield insights into the dynamic behavior of the disease spread. With new insights, health officials can develop more effective disease intervention strategies. Besides, such epidemic models are also used to forecast the course of the epidemic.

In addition to epidemic models, we consider the statistical machine learning (SML) models, which are extremely good for prediction. Often, the interpretability of SML models is questioned. However, as we take a model agnostic approach; we can use the epidemic models to understand the ground reality while adopting the SML to achieve better prediction accuracy.

\subsection{SIR Epidemiological Model}

The popular epidemic models for an infectious disease is the Susceptible, Infected, Recovered (SIR) model. The model considers a closed population. To start with, a few infected people are added to the population. It assumes that the mixing pattern is homogeneous. During the period of the sickness, the contagious people each infect on average $\mathcal{R}_0$ other people, who each then go on to infect $\mathcal{R}_0$ others, who are susceptible. The $\mathcal{R}_0$ is popularly known as the Basic Reproduction Number. The $\mathcal{R}_0$ is the fundamental quantity of the disease progression, and higher $\mathcal{R}_0$ means, more people will tend to be infected in the course of the epidemic. The major advantage of the SIR model is it gives a  number $\mathcal{R}_0$, which can be used to benchmark and compare the ground situation of different states and resource allocations can be made to those states which are hard hit. The SIR model can be described as,
\begin{eqnarray}
\label{SIR_model_eqn}
\frac{\partial S}{\partial t}&=&-\beta \frac{SI}{N}\nonumber\\
\frac{\partial I}{\partial t}&=&+\beta \frac{SI}{N}-\gamma I\\
\frac{\partial R}{\partial t}&=&+\gamma I\nonumber
\end{eqnarray}
where $S$, $I$, and $R$ are the number of people in the population that are susceptible, infected and recovered. The $\beta$ is the transmission rate. Each susceptible person contacts $\beta$ people per day; a fraction $\frac{I}{N}$ of which are infectious. Therefore $\beta \frac{SI}{N}$ move out of the susceptible group and goes into the infected group. The transmission rate is the average rate of contacts a susceptible person makes that is sufficient to transmit the infection.  The parameter $\gamma$ is the recovery rate, and $\gamma I$ is the flow out of the infected crowd and goes into the recovered group. The average duration a person spends in the infected group is $\frac{1}{\gamma}$ days. For Covid-19, $\frac{1}{\gamma}$ is around 14 days, see \cite{ferguson2020}. 

In this paper, we follow the SIR implementation methodology as described in \cite{Sherry_Towers_SIR_model}. Given $\mathcal{R}_0$, $\beta$ and $\gamma$, the implementation of SIR model is fairly straight forward via \texttt{deSolve} package, a solvers for initial value problems of differential equations, see \cite{deSolve_package}. It is known that $\mathcal{R}_0=\frac{\beta}{\gamma}$, see \cite{Math_Epi}. We considered $\gamma$ as $\frac{1}{14}$, from \cite{ferguson2020}. However, we need some good estimates of the $\mathcal{R}_0$, so that we can implement the SIR model and predict the disease progression in India. In order to estimate the $\mathcal{R}_0$, we use the \texttt{R}-package called, `\texttt{R0}', a toolbox to estimate $\mathcal{R}_0$, see \cite{R0_obadia}. The time between the infection of a primary case and one of its secondary cases is called a generation time, see \cite{sevensson2007}. The `\texttt{R0}' package assumes generation time of infection is known and should be provided as input. The mean generation time for the Wuhan has been reported as 6.5 days, \cite{Wuhan_study_NEJM_2020}. In this paper, we assume the generation time follows Gamma distribution and we estimated the mean and shape parameter of the Gamma distribution using data. Our estimated mean generation time for Hubei province turns out to be 6.7, presented in the Table \ref{table_China_India_R0}. On the recovery from infection, we assume the individuals are assumed to be immune to re-infection in the short term. This assumption is same as \cite{ferguson2020}. 

Currently, we are deploying a grid search method over the mean and shape of the Gamma distribution for the time generation process. For a particular choice of the mean ($\mu$) and shape ($\kappa$) parameter, we generate the time and then given that as input we estimate the $\mathcal{R}_0$ using the `\texttt{R0}' package in \texttt{R}. Then for an estimated $\mathcal{R}_0$ and $\gamma$ (assumed to be 1/14), we simulate the disease progression, for the period, for which we observed the new incidences. Then we calculate the Mean Square Error (MSE) in the following way:
\begin{equation}
\label{Eqn_MSE}
MSE(\mu,\kappa)=\frac{1}{T}\sum_{t=1}^T\Big(\hat{I}(t)-i_{obs}(t)\Big)^2,
\end{equation}
where $\hat{I}(t)$ is the new incidence estimated from SIR model described in (\ref{SIR_model_eqn}) at time point $t$, and $i_{obs}(t)$ is the actual incidence observed in the data at time point $t$. We estimate the mean parameter $\mu$ and shape parameter $\kappa$ for which the MSE in (\ref{Eqn_MSE}) is minimum. The, for estimated mean and shape parameter, $\mathcal{R}_0$ is estimated using the `\texttt{R0}' package. 

\subsection{Statistical Machine Learning Model}

The infection rate of a typical epidemic reaches its peak and then it slows down. The SIR model predicts when that peak will be reached very well because it captures the inherent dynamism of the epidemic. However, the SIR model is not helpful for short and medium-term predictions. We also need short and medium-term prediction, to predict the cases as quickly as possible so that the health officials can take the appropriate decision. The Statistical Machine Learning (SML) models are most popular for its prediction accuracy from short to medium term, \cite{Sambasivan2020}. Consequently, SML and SIR models complement each other. Note that the SML does not do well in long term prediction, particularly it cannot predict when it will reach the peak. Under this understanding, we develop traditional SML models and not deep learning models. We refrain to develop deep learning type models because we need a lot of data. However, in epidemiology, we do not have such kind of big data. In addition, the literature on how to adopt deep-learning for small data is not sufficient yet. Therefore we refrain from developing deep learning models and we develop the traditional regression type SML model, for short to medium type prediction.

As different countries or provinces population levels are different; we consider the our variable to analyze as cases per 100,000 (aka., Rate),
$$
\text{Rate} =\frac{\text{Cases}}{\text{Population Size}}\times 100,000.
$$
The we model the Rate as a function of time, country and time-country interaction in the following way:
\begin{eqnarray}
\ln\{\text{Rate}_{it}+1\}&=&\beta_0+\beta_1 t+\beta_2 t^2+\cdots+\beta_p t^p \nonumber\\
&& +\alpha_i + \alpha_i t+ \alpha_i t^2 +\cdots +\alpha_i t^p +\epsilon, \label{eqn_SML_prediction_model}
\end{eqnarray}
where $\text{Rate}_{it}$ is the Rate of the $i^{th}$ country at the $t^{th}$ time point, $\alpha_i$ is the effect of $i^{th}$ country, $\alpha_i t$ is the linear effect of time on the Rate of the $i^{th}$ country, $\alpha_i t^2$  is the quadratic effect of time on the Rate of the $i^{th}$ country. We considered the following countries in our model: (1) India, (2) China, (3) US, (4) Iran, (5) South Korea, (5) Japan, (6) Italy, (7) France, (8) Germany, and (9) Spain. 

\subsection{Model Training Strategy for India to Measure the Effect of the Lockdown}

On March 24, 2020, India announced the national lockdown of the nation. To measure the effectiveness of the lockdown, we used all data up to March 24, 2020, to train the model and learn the parameters of the model. Based on the trained model, we predict the disease progression path. Since the incubation period of the  COVID-19 is about 14 days, it is likely that for 14 days from the beginning of the lockdown, the disease will follow the predicted path and then, it will deviate down from the predicted path. If the new confirmed cases come below the predicted path then we will know that is due to the effect of lockdown. On the other hand, if the disease progression stays on the predicted path then we will know the lockdown did not work. If the disease progression comes above the predicted path then we can say that the ground situation  
worsen during the lockdown.

\section{Analysis and Prediction}\label{Sec_Analysis_and_Prediction}

\begin{figure}[ht]
    \centering
    \includegraphics[height=10cm]{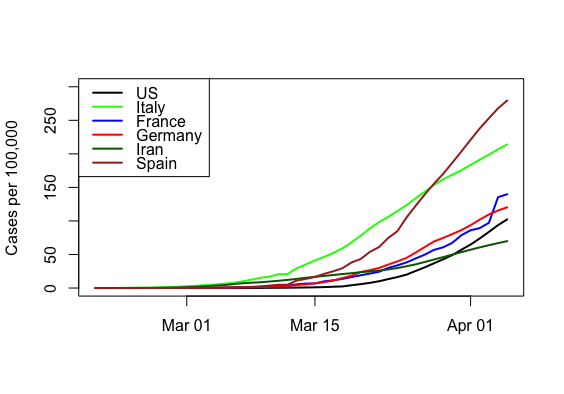}
    \caption{Cases per 100,000 in the US, EU and, Iran. The worst hit US, EU and Iran's cases per 100,000 is in the range of 70 to 250. All countries are }
    \label{fig_cases_per_100000_US_EU_Iran}
\end{figure}

\begin{figure}[ht]
    \centering
    \begin{tabular}{cc}
      \includegraphics[height=7cm]{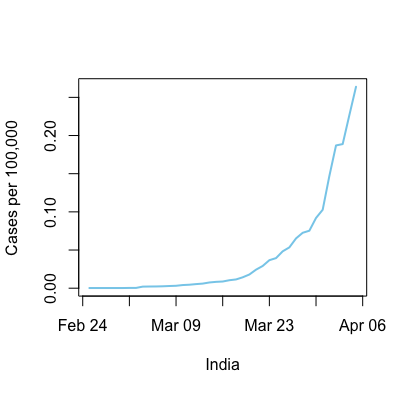}   &  \includegraphics[height=7cm]{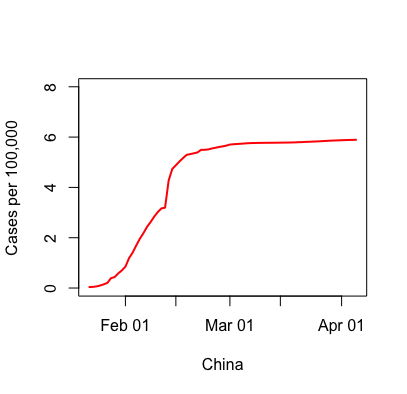} \\
      \includegraphics[height=7cm]{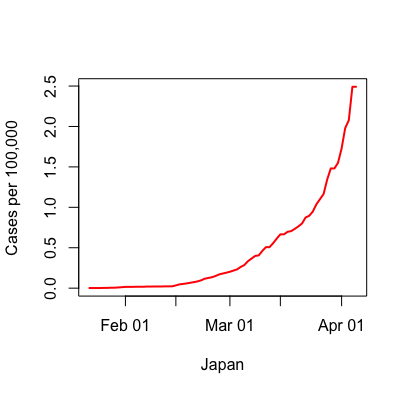}     &  \includegraphics[height=7cm]{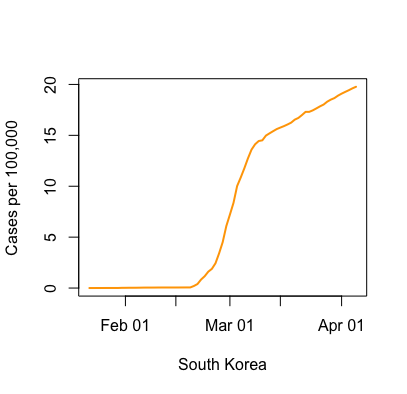}   
    \end{tabular}
    \caption{Cases per 100,000 in India, China, Japan, and South Korea. Note that India and Japan's cases per 100,000 are in exponential rise. However, China and South Korea were able to flatten the curve. But at different levels. China was able to flatten the curve at around 6 per 100,000 population; whereas South Korea has partially flattened its curve and increasing as a linear scale.}
    \label{fig_cases_per_100000_Asia}
\end{figure}

\paragraph{Exploratory Data Analysis} (EDA) is important to develop good predictive models. In the Figure (\ref{fig_cases_per_100000_US_EU_Iran}), we plot the case per 100,000 (aka., Rate) for US, EU and Iran. The worst-hit US, EU and Iran's rates are in the range of 70 and  250. On the other hand, disease progression among Asian countries is very different, see Figure (\ref{fig_cases_per_100000_Asia}). The disease progression for both India and Japan are similar. We see the exponential rise in India and Japan but at a very lower rate than the Western nations. China has able to flatten the curve and South Korea was able to curb the rise from exponential to linear. However, so far South Korea experienced the worst rate among the major four Asian countries.

\begin{figure}[ht]
    \centering
    \includegraphics[height=12cm]{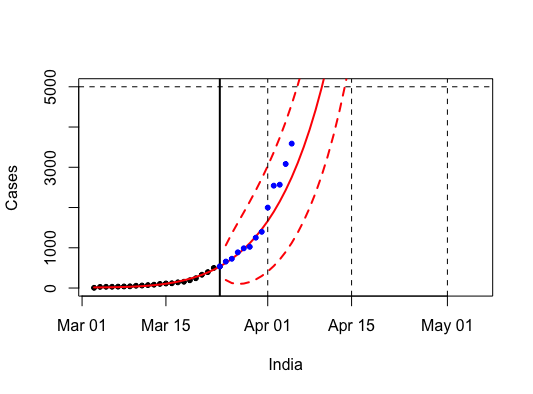}
    \caption{Predicted path of the disease progression in India.  The solid black vertical line represent the 24 March 2020. The black points left of the vertical black line are confirmed cases till 24 March 2020. These black points are used in model training. The solid red line is the predicted path of the disease progression. The blue points are the out of sample test point or the confirmed cases that comes after 24 March 2020. As of 07 April 2020, we don't see the effect of lockdown. However, if lockdown works - it should shows its effect any time soon now. The blue point should appear below the predicted red line.}
    \label{fig_india_prediction_20200324}
\end{figure}

\paragraph{Prediction of Disease Progression for India} from SML model (\ref{eqn_SML_prediction_model}). The solid black vertical line in the Figure (\ref{fig_india_prediction_20200324}) represent the 24 March 2020. The black points left of the vertical black line are confirmed cases till 24 March 2020. These black points are used in model training. The solid red line is the predicted path of the disease progression. The blue points are the out of sample test point or the confirmed cases that comes after 24 March 2020. As of 07 April 2020, we don't see the effect of lockdown. However, if lockdown works - it should shows its effect any time soon now. The blue point should appear below the predicted red line. In the Table (\ref{tabl_India_prediction}), we present the actual prediction till May 01, 2020. If lockdown works then actual confirmed cases for India should stay below 66,224 by May 01,2020.

\paragraph{A Comparison of $\mathcal{R}_0$ between India and China}: In the Table (\ref{table_China_India_R0}), $\mathcal{R}_0$ with a 95\% confidence interval for Hubei province and China is around 2.5 during the first 23 days from the starting of the Lockdown. India's $\mathcal{R}_0$ with a 95\% confidence interval computed using two different starting points as breakout. One from 02-Mar-2020, because the number of cases in India started rising from that day. The $\mathcal{R}_0$ for India for the first 22 days till the lockdown is around 2.5, like China. However, if we use the data, till 04-Apr-2020, then the $\mathcal{R}_0$ value is around 2.75. It indicates since the lockdown the situation has worsen. It is also clear from the Figure (\ref{fig_india_prediction_20200324}). In the second approach, we consider India's breakout from 23-Jan-2020. In that situation, if we consider the data till 24-Mar-2020, the $\mathcal{R}_0$ with 95\% confidence is almost 1.9 and if we consider data till 04-Apr-2020, the $\mathcal{R}_0$ is nearly 2.1. It means if we use the data earlier to 02-Mar-2020 the India's $\mathcal{R}_0$ looks better. 

\begin{figure}[ht]
    \centering
    \begin{tabular}{cc}
    \includegraphics[height=7cm]{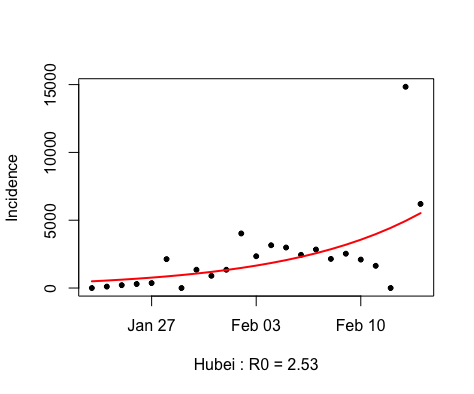}
         &  
    \includegraphics[height=7cm]{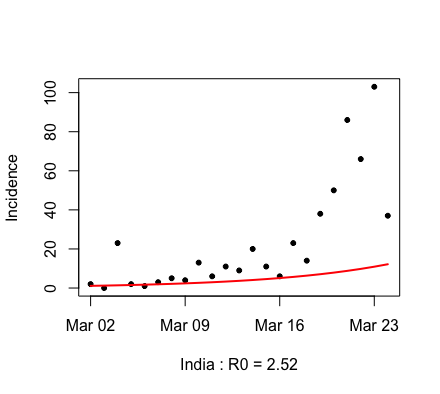}
    \\
    (a) &(b)
    \end{tabular}
    \caption{In this figure, we compare the incidences of Hubei and India in (a) and (b). We consider the date range for Hubei from 23-Jan-2020 to 14-Feb-2020, i.e., during the first 23 days of Hubei lockdown. On the other hand, we considered the data for India, from the 02-Jan-2020 to 24-Jan-2020, before the lockdown. On the 23-Jan-2020, Hubei had 444 confirmed cases and overall China had 548 confirmed cases. On 02-Jan-2020, India had only 3 confirmed cases, whereas on the day of lockdown, i.e., on 24-Jan-2020, India had 536 confirmed cases.}
    \label{fig_Hubei_India}
\end{figure}

In the Figure (\ref{fig_Hubei_India}), we compare the incidences of Hubei and India in Figure (\ref{fig_Hubei_India}:a) and (\ref{fig_Hubei_India}:b). We consider the date range for Hubei from 23-Jan-2020 to 14-Feb-2020, i.e., during the first 23 days of Hubei lockdown. On the other hand, we considered the data for India, from the 02-Jan-2020 to 24-Jan-2020, till the lockdown. On the 23-Jan-2020, Hubei had 444 confirmed cases and overall China had 548 confirmed cases. On 02-Jan-2020, India had only 3 confirmed cases, whereas on the day of lockdown, i.e., on 24-Jan-2020, India had 536 confirmed cases. So on the day, when the lockdown starts both India and Hubei and/or China had a comparable number of cases. Perhaps, we should consider India's $\mathcal{R}_0$ as around 2.5 similar to that of the early stage COVID-19 disease progression of China. Even with the lockdown, China experienced more than 80,000 cases. Perhaps, we should prepare for at least that many cases if not more in India.

\begin{figure}[ht]
    \centering
    \includegraphics[height=10cm]{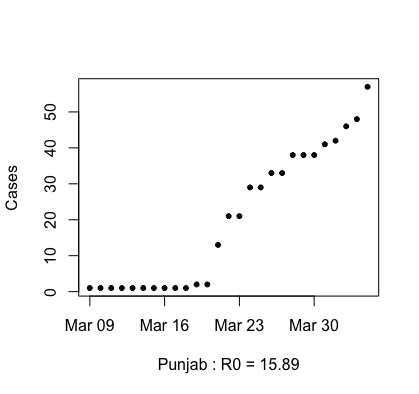}
    \caption{Confirmed cases of COVID19 in Punjab. The $\mathcal{R}_0=15.89$. The high $\mathcal{R}_0$ is likely due to a super spreader ignored advice to self quarantine after returning from a trip to Italy and Germany, see \cite{BBC_report_India_super_spreader}}.
    \label{Fig_Punjab_cases}
\end{figure}

\paragraph{Stat wise $\mathcal{R}_0$}: In Table (\ref{Table_statewise_R0}), we present the state wise Basic Reproduction Number, $\mathcal{R}_0$, as of 04 March, 2020. We see the Punjab's $\mathcal{R}_0$ is worst in the country. Punjab's high $\mathcal{R}_0\approx 16$ is likely due to a super spreader, who ignored advice to self quarantine after returning from a trip to Italy and Germany, see \cite{BBC_report_India_super_spreader}. The situation is Punjab is really complicated and serious intervention is required. In Figure (\ref{Fig_Punjab_cases}), we present the cases in Punjab over time. Since March 20, 2020 the number of confirmed cases increased at an unprecedented rate. From Table (\ref{Table_statewise_R0}), we see the $\mathcal{R}_0$ for Madhya Pradesh (3.37) , Maharastra (3.25) and Tamil Nadu (3.09) are more than 3. Clearly the situations are complicated in these three states. The $\mathcal{R}_0$ of Andhra Pradesh (2.96), Delhi (2.82) and West Bengal (2.77) is more than the India's $\mathcal{R}_0$ which is 2.75. These seven states should need special attention as their $\mathcal{R}_0$ is more than that of India (2.75). These numbers are as of 4 Apr, 2020. For the following states, we either do not have enough data to make inference for $\mathcal{R}_0$; or the algorithm fail to converge: (1) Andaman and Nicobar Islands; (2) Arunachal Pradesh; (3) Chattisgarh; (4) Goa; (5) Haryana; (6) Jharkhand; (7) Manipur; (8) Mizoram; (9) Odisha; (10) Puducherry.

\newpage 

\section{Discussion}\label{Sec_Discussion}
Here we present a point by point discussion of our analysis and prediction.
\begin{enumerate}
    \item Situation of Punjab ($\mathcal{R}_0\approx 16$) is bad. It requires immediate aggressive attention.
    \item We see the $\mathcal{R}_0$ for Madhya Pradesh (3.37) , Maharastra (3.25) and Tamil Nadu (3.09) are more than 3. Aggressive intervention is needed.
    \item The $\mathcal{R}_0$ of Andhra Pradesh (2.96), Delhi (2.82) and West Bengal (2.77) is more than the India's $\mathcal{R}_0=2.75$.
    \item India's $\mathcal{R}_0=2.75$ (as of 04 March, 2020) is very much comparable to Hubei/China at the early disease progression stage.
    \item Our analysis indicates that the early disease progression of India is that of similar to China. Therefore, with lockdown in place, India should expect as many as cases if not more like China.
    \item If lockdown works, we should expect less than 66,224 cases by May 01,2020.
\end{enumerate}

\bibliographystyle{abbrv}
\bibliography{bibliography_database}

\newpage

\section*{Tables and Figures}

\begin{table}[ht]
\centering
\begin{tabular}{lccccccc}
  \hline
 & & &  &  & Initial Number of &  & \\ 
 &Date Range & $\mathcal{R}_0$ & $\mathcal{R}_0$ Lower& $\mathcal{R}_0$ Upper&infections considered & mean  & shape \\
  &&&&& in SIR model&($\hat{\mu}$)&($\kappa$) \\
  \hline
Hubei & 23-Jan-20 to 14-Feb-20  & 2.53 & 2.50 & 2.57 & 444& 6.7 & 0.24 \\ 
  China & 23-Jan-20 to 14-Feb-20  & 2.46 & 2.43 & 2.49 & 548& 8.7 & 2.7\\ \hline
  India & 02-Mar-20 to 24-Mar-20 & 2.52 & 2.35 & 2.71 & 3& 5.84 & 6.56\\ 
  India & 02-Mar-20 to 04-Apr-20 & 2.75 & 2.63 & 2.89 & 3& 5.41 & 1.10\\ \hline
  India & 23-Jan-20 to 24-Mar-20 & 1.87 & 1.78 & 1.97 & 1& 2.96 & 1.53\\ 
  India & 23-Jan-20 to 04-Apr-20 & 2.09 & 2.04 & 2.14 & 1& 1.25 & 4.98\\ 
   \hline
\end{tabular}
\caption{$\mathcal{R}_0$ with a 95\% confidence interval for Hubei province and China is around 2.5 during the first 23 days from the starting of the Lockdown. India's $\mathcal{R}_0$ with a 95\% confidence interval using two different starting points. One from 02-Mar-2020, because the number of cases in India started rising from that day. The $\mathcal{R}_0$ for India for the first 22 days till the lockdown is around 2.5, like China. However, if we use the data, till 04-Apr-2020, then the $\mathcal{R}_0$ value is around 2.75. In the second approach, we consider India's breakout from 23-Jan-2020. In that situation, if we consider the data till 24-Mar-2020, the $\mathcal{R}_0$ with 95\% confidence is almost 1.9 and if we consider data till 04-Apr-2020, the $\mathcal{R}_0$ is nearly 2.1.}
\label{table_China_India_R0}
\end{table}

\begin{table}[ht]
\centering
\begin{tabular}{lrrr}
  \hline
  State/UT & $\mathcal{R}_0$ & Lower & Upper \\ 
  \hline
 Andhra Pradesh & 2.96 & 2.56 & 3.45 \\ 
 Bihar & 2.13 & 1.35 & 3.40 \\ 
 Chandigarh & 1.14 & 0.89 & 1.48 \\ 
 Delhi & 2.82 & 2.60 & 3.08 \\ 
 Gujarat & 0.98 & 0.84 & 1.15 \\ 
 Himachal Pradesh & 1.59 & 1.00 & 3.13 \\ 
 Jammu and Kashmir & 2.02 & 1.69 & 2.48 \\ 
 Karnataka & 2.29 & 1.87 & 2.77 \\ 
 Kerala & 1.62 & 1.52 & 1.74 \\ 
 Ladakh & 1.54 & 1.17 & 2.18 \\ 
 Madhya Pradesh & 3.37 & 2.73 & 4.14 \\ 
 Maharashtra & 3.25 & 2.95 & 3.58 \\ 
 \textbf{Punjab} & \textbf{15.89} & \textbf{4.12} & \textbf{149.27} \\ 
 Rajasthan & 2.45 & 2.25 & 2.67 \\ 
 Tamil Nadu & 3.09 & 2.74 & 3.53 \\ 
 Telengana & 2.16 & 1.97 & 2.38 \\ 
 Uttar Pradesh & 2.30 & 2.10 & 2.52 \\ 
 Uttarakhand & 1.33 & 1.13 & 1.61 \\ 
 West Bengal & 2.77 & 2.21 & 3.47 \\ \hline
 India & 2.75 & 2.63 & 2.89 \\
 \hline
\end{tabular}
\caption{State Wise Basic Reproduction Number, $\mathcal{R}_0$, as of 04 March, 2020. Punjab's high $\mathcal{R}_0$ is likely due to a super spreader ignored advice to self quarantine after returning from a trip to Italy and Germany, see \cite{BBC_report_India_super_spreader}}.
\label{Table_statewise_R0}
\end{table}

\begin{table}[ht]
\centering
\begin{tabular}{rlrr}
  \hline
 & Dates & Actual Case & Prediction \\ 
  \hline
1 & 2020-03-03 &   5 & 14.99 \\ 
  5 & 2020-03-07 &  34 & 22.42 \\ 
  10 & 2020-03-12 &  73 & 57.72 \\ 
  15 & 2020-03-17 & 142 & 158.74 \\ 
  20 & 2020-03-22 & 396 & 387.54 \\ 
  21 & 2020-03-23 & 499 & 456.29 \\ 
  22 & 2020-03-24 & 536 & 534.79 \\ \hline
  23 & 2020-03-25 & 657 & 624.10 \\ 
  24 & 2020-03-26 & 727 & 725.36 \\ 
  25 & 2020-03-27 & 887 & 839.85 \\ 
  26 & 2020-03-28 & 987 & 968.95 \\ 
  27 & 2020-03-29 & 1024 & 1114.20 \\ 
  28 & 2020-03-30 & 1251 & 1277.28 \\ 
  29 & 2020-03-31 & 1397 & 1460.05 \\ 
  30 & 2020-04-01 & 1998 & 1664.59 \\ 
  31 & 2020-04-02 & 2543 & 1893.20 \\ 
  32 & 2020-04-03 & 2567 & 2148.44 \\ 
  33 & 2020-04-04 & 3082 & 2433.18 \\ 
  34 & 2020-04-05 & 3588 & 2750.66 \\ 
  35 & 2020-04-06 & 4778 & 3104.50 \\ 
  39 & 2020-04-10 &  & 4974.57 \\ 
  44 & 2020-04-15 &  & 8838.36 \\ 
  49 & 2020-04-20 &  & 15791.88 \\ 
  54 & 2020-04-25 &  & 29126.81 \\ 
  59 & 2020-04-30 &  & 57229.81 \\ 
  60 & 2020-05-01 &  & 66223.94 \\ 
   \hline
\end{tabular}
\caption{The table presents the actual cases and prediction from the SML model (\ref{eqn_SML_prediction_model}). We used all the data till the 24th March 2020. Here due to space constraint, we present only 5 days interval and recent out of sample values at the daily level.}
\label{tabl_India_prediction}
\end{table}

\end{document}